\documentclass[3p,times,procedia]{elsarticle}
\usepackage{nupha_ecrc}

\volume{00}

\firstpage{1}

\journalname{Nuclear Physics A}

\runauth{E.~Iancu et al.}

\jid{nupha}

\jnltitlelogo{Nuclear Physics A}

\usepackage{bm,amsmath,amssymb}

\usepackage[figuresright]{rotating}

\long\def\comment#1{ }
\newcommand{\eqn}[1]{Eq.~\eqref{#1}}
\newcommand{\beq}{\begin{equation}}
\newcommand{\eeq}{\end{equation}}

\newcommand{\rmd}{{\rm d}}
\newcommand{\rme}{{\rm e}}
\newcommand{\rmi}{{\rm i}}

\newcommand{\bk}{\bm{k}}

\newcommand{\bx}{\bm{x}}
\newcommand{\by}{\bm{y}}

\newcommand{\bz}{\bm{z}}

\newcommand{\br}{\bm{r}}

\newcommand{\kt}{k_\perp} 

\newcommand{\abar}{\bar{\alpha}_s}
\newcommand{\order}[1]{\mcal{O}{(#1)}}
\newcommand{\mcal}{\mathcal}

\newcommand{\nlo}{{\rm \scriptscriptstyle NLO}}
\newcommand{\lo}{{\rm \scriptscriptstyle LO}}

\newcommand{\CXY}{{\rm \scriptscriptstyle CXY}}

\newcommand{\calS}{\mathcal{S}}

\newcommand{\calK}{\mathcal{K}}
\newcommand{\minus}{\!-\!}

\begin{document}

\begin{frontmatter}



\dochead{XXVIth International Conference on Ultrarelativistic Nucleus-Nucleus Collisions\\ (Quark Matter 2017)}

\title{Particle production in $pA$ collisions beyond leading order}

\author[sac]{E.~Iancu}

\author[col]{A.H.~Mueller}

\author[ect]{and D.N.~Triantafyllopoulos}

\address[sac]{Institut de physique th\'{e}orique, Universit\'{e} Paris Saclay, CNRS, CEA, F-91191 Gif-sur-Yvette, France}

\address[col]{Department of Physics, Columbia University, New York, NY 10027, USA}

\address[ect]{ECT* and Fondazione Bruno Kessler, Strada delle Tabarelle 286, I-38123 Villazzano (TN), Italy}



\begin{abstract}
We describe the origin of, and the solution to, the negativity problem which occurs in the perturbative calculation
of the cross-section for single-inclusive particle production in proton-nucleus collisions at next-to-leading-order.
\end{abstract}


\end{frontmatter}


\section{Introduction}
\label{sec:0}

Particle production at forward rapidities and semi-hard transverse momenta in proton (or deuteron)--nucleus
collisions at RHIC and the LHC is an important source of information about the small-$x$ part of the nuclear
wavefunction, where gluon occupation numbers are high and non-linear effects like gluon saturation and
multiple scattering are expected to be important. On the theory side, the cross-section 
for single-inclusive particle production has been computed \cite{Chirilli:2012jd} 
 up to next-to-leading order (NLO) in the framework of the so-called `hybrid factorization' \cite{Dumitru:2005gt}, 
but the result is problematic: the cross-section suddenly turns negative when increasing the transverse
momentum of the produced hadron, while still in the semi-hard regime \cite{Stasto:2013cha}.
Various proposals to fix this difficulty, by modifying the scale for the rapidity subtraction, have
only managed to push the problem to somewhat larger values of the 
transverse momentum (see \cite{Stasto:2016wrf} for a recent discussion and more references).
In a recent paper \cite{Iancu:2016vyg}, we have argued that this negativity problem is an
artifact of some of the approximations inherent in hybrid factorization, as needed to obtain
a result which looks local in rapidity. On that occasion, we have also proposed a more general
factorization scheme (within the CGC effective theory \cite{Gelis:2010nm}), which
is non-local in rapidity but yields a manifestly positive cross-section to NLO accuracy.


\section{Leading order formalism}
\label{sec:1}

To leading order (LO) in the CGC effective theory, quark production at forward rapidities in $pA$ collisions proceeds
as follows: a quark from the wavefunction of the incoming proton, which is initially collinear with the
proton and carries a relatively large longitudinal fraction $x_p$, scatters off the dense gluon distribution
in the nuclear target and thus acquires a transverse momentum $\bk$. The 
quark distribution is computed as
\beq\label{LO}
 \frac{\rmd N^{pA\to qX}}{\rmd^2\bk\, \rmd \eta}\bigg|_{\lo}
=\frac{1}{(2\pi)^2}\,
 x_p q(x_p)\,
{\calS}(\bk,X_g)\,,\qquad {\calS}(\bk,X_g)=\int \rmd^2\br\, \rme^{-\rmi \bk \cdot \br} {S}(\br,X_g),
\eeq
where $\eta$ is the rapidity of the produced quark in the center-of-mass frame and 
$X_g$ is longitudinal momentum
fraction carried by the gluons from the target that are involved in the collision.
Energy-momentum conservation implies
$x_p=({\kt}/\sqrt{s})\rme^\eta$ and $X_g=({\kt}/\sqrt{s})\rme^{-\eta}$.
The {\em forward kinematics} corresponds to the situation where $\eta$ 
is positive and large, which implies $X_g\!\ll\! x_p \!< \! 1$. Thus,
forward particle production explores the small-$X_g$ part of the nuclear wavefunction,
as anticipated. Furthermore, $x_p q(x_p)$ is the quark distribution of the proton and
${S}(\br,X_g)$ is the elastic $S$--matrix for the scattering
between a `color dipole' (a quark-antiquark pair in a color-singlet state) and the nucleus.
Its Fourier transform  $\calS(\bk,X_g)$ plays the role of an
unintegrated gluon distribution in the nuclear target, as probed by particle production
in dilute-dense collisions.

 \begin{figure}[t]
\centerline{
\includegraphics[width=.75\textwidth]{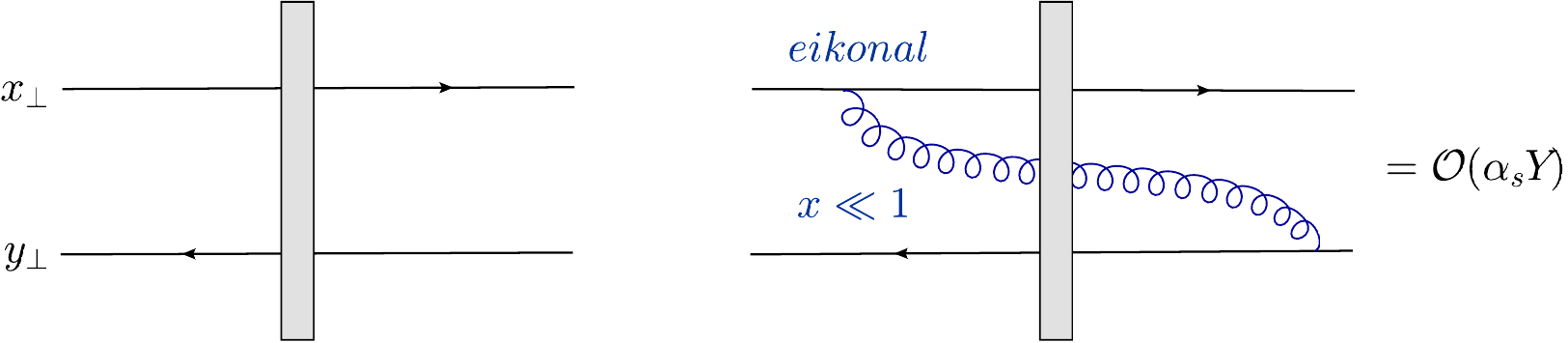}}
\caption{\sl Leading-order contributions to the dipole $S$-matrix.
Left: the scattering of the bare dipole, which yields 
$S_0$. Right: one step in the LO evolution (the emission of a soft gluon
in the eikonal approximation),  yielding a contribution of $\order{\alpha_s Y}$
with $Y=\ln(1/X_g)$}
\label{fig:eik}
\end{figure}

To the same accuracy, the dipole $S$-matrix is obtained by solving 
the LO version of the BK equation \cite{Gelis:2010nm}, 
which resums an arbitrary number of soft gluon emissions
in the scattering between the dipole and the nuclear target, in the eikonal approximation. 
For what follows, it is convenient to chose a Lorentz frame in which 
the `primary' gluon (the one which is closest in rapidity to the dipole) is emitted by the dipole,
whereas all the other, even softer, gluons belong to the nuclear wavefunction.
Then, the LO dipole $S$-matrix ${S}(\br,X_g)\equiv S\big(\bx,\by; X_g\big)$  (with $\br=\bx-\by$)
admits the following integral representation (see also Fig.~\ref{fig:eik})
\begin{align}\label{BKint}
 S\big(\bx,\by; X_g\big)=S_0(\bx,\by)+
 \frac{\abar}{2\pi} \int_{X_g/X_0}^{1}\frac{\rmd x}{x}\int 
\frac{\rmd^2\bz \,(\bx-\by)^2}{(\bx-\bz)^2(\bz-\by)^2}
\Big[
 S\big(\bx,\bz; X(x)\big) S\big(\bz, \by; X(x)\big)
 -S\big(\bx,\by; X(x)\big)\Big]\,,
 \end{align}
where $\abar=\alpha_s N_c/\pi$,
$\bx$ and $\by$ are the transverse coordinates of the quark and the antiquark
(which are not changed by an eikonal emission), 
$\bz$ is the transverse position of the primary gluon,
$X_0$ is the value of $X$ at which one starts the high-energy evolution 
of the target, $S_0=S(X_0)$ is the corresponding initial condition (say, as given by
the McLerran-Venugopalan model), $x\ll 1$ is the fraction of the dipole longitudinal momentum
taken by the primary gluon, and $X(x)\equiv X_g/x$ is the longitudinal 
momentum fraction of the gluons from the target which scatter off the projectile made with the dipole
and the primary gluon. 

\section{Beyond leading order}
\label{sec:2}

\begin{figure}[t]
\centerline{
\includegraphics[width=.75\textwidth]{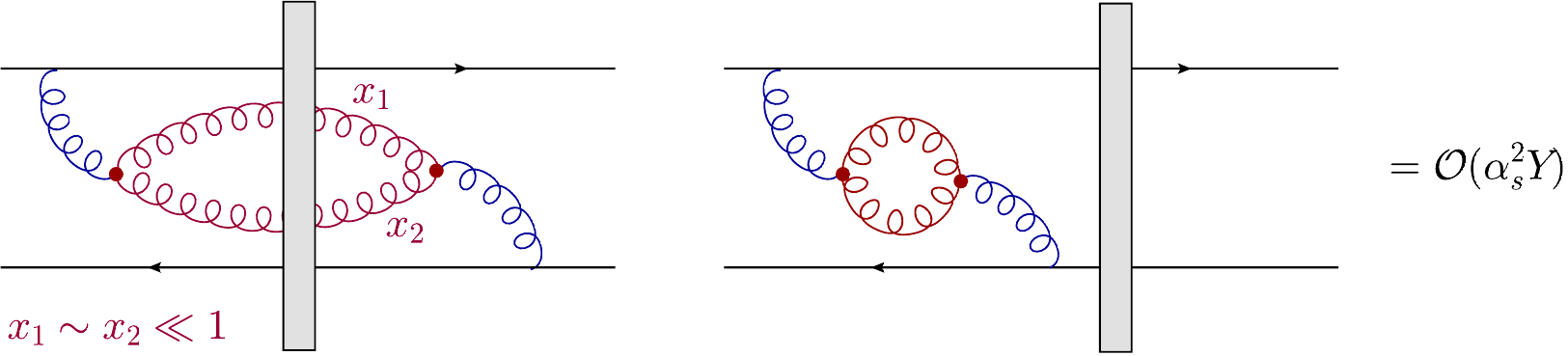}}
\caption{\sl Next-to-leading order correction to the kernel of the BK equation}
\label{fig:evol}
\end{figure}

\begin{figure}[t]
\centerline{
\includegraphics[width=.75\textwidth]{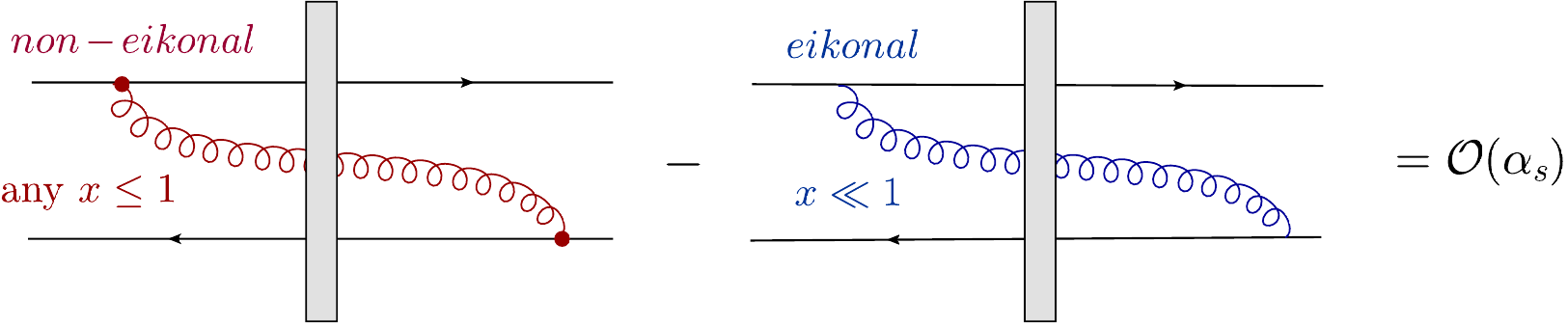}}
\caption{\sl Next-to-leading order correction to the quark impact factor in the dipole picture.}
\label{fig:IF}
\end{figure}

The quark multiplicity in \eqn{LO} receives two types of next-to-leading order (NLO) corrections
(see Figs.~\ref{fig:evol} and \ref{fig:IF}):
those related to the dipole evolution --- i.e. corrections of $\order{\alpha_s}$ to the kernel
of the BK equation \cite{Balitsky:2008zza} ---
and those related to the impact factor --- corrections of $\order{\alpha_s}$ which arise
when the emission of the primary gluon is computed ``with exact kinematics'', 
i.e. beyond the eikonal approximation \cite{Chirilli:2012jd}. 

The NLO corrections to the BK kernel are generated by the ensemble of the genuine one-loop corrections 
to the emission of a soft ($x\ll 1$) primary gluon. For instance, Fig.~\ref{fig:evol} illustrates the effect of a gluon 
loop where the secondary gluons are both soft, but their longitudinal momentum fractions
are comparable to each other: $x_1\sim x_2 \ll 1$.  (The situation where $x_2\ll x\simeq x_1$ contributes to the 
second step of the LO evolution and must be properly subtracted when computed the NLO correction
to the evolution kernel.) Accordingly, the 3-gluon vertices visible in Fig.~\ref{fig:evol} must be computed
with exact kinematics, and similarly for the other one-loop diagrams.
The ensemble of such corrections has been computed in \cite{Balitsky:2008zza}.
The strict NLO version of the BK equation turns out to be unstable
\cite{Lappi:2015fma}, due to the presence of large
NLO corrections enhanced by collinear logarithms. All-order resummations which overcome this
problem have been devised in \cite{Beuf:2014uia,Iancu:2015vea,Iancu:2015joa}.

In what follows we shall focus on the NLO correction to the impact factor, which as we shall see
is responsible for the negativity problem mentioned in the introduction. To isolate this correction,
one must compute the primary emission with the exact kinematics (as valid for any $x\le 1$)
and subtract away the respective contribution of an eikonal emission (strictly correct for $x\ll 1$ alone), 
that was already included in the LO BK evolution. This is illustrated in Fig.~\ref{fig:IF}. To simplify 
the discussion, we shall keep the dipole evolution at LO (possibly amended by running coupling corrections;
see below). Also, we shall not present the NLO corrections in detail (these can be found in the literature;
see e.g. \cite{Iancu:2016vyg,Ducloue:2017mpb}),
rather we shall use symbolic notations, which are both compact and suggestive.


To start with, we shall rewrite the LO BK equation \eqref{BKint} as [we recall
that $\br=\bx-\by$ and $X(x)=X_g/x$]
\beq\label{toyLOBK}
{S}(\br, X_g)=S_0(\br) + \abar \int_{X_g/X_0}^{1} \frac{\rmd x}{x} \,K(\br; 0) \,
S\big(\br, X(x)\big)\,,\eeq
where the kernel $K(\br; 0)$ succinctly denotes the convolution in transverse space
and the terms quadratic in $S$ are kept implicit. Clearly, a similar representation can be 
written in momentum space, with $K(\br; 0)\to \calK(\bk; 0)$. In these symbolic
notations, the LO cross-section \eqref{LO} reads
\beq\label{toyLON}
 \frac{\rmd N^{pA\to qX}}{\rmd^2\bk\, \rmd \eta}\bigg|_{\lo}
=\,{\calS}(\bk,X_g)\,=\,
\calS_0(\bk) + \abar\int_{X_g/X_0}^{1} \frac{\rmd x}{x} \,\calK(\bk; 0) \,
\calS\big(\bk, X(x)\big)\,,\eeq
where we also have omitted the quark distribution, to simplify notations. Notice that the last
expression in  \eqref{toyLON} is the integral representation for ${\calS}(\bk,X_g)$ obtained
by taking a Fourier transform in \eqref{toyLOBK}.

This last expression can be generalized to include the NLO correction to the impact factor  \cite{Iancu:2016vyg}:
\beq\label{toyNLON}
 \frac{\rmd N^{pA\to qX}}{\rmd^2\bk\, \rmd \eta}\bigg|_{\nlo}=\,
\calS_0(\bk) + \abar\int_{X_g/X_0}^{1} \frac{\rmd x}{x} \,\calK(\bk; x) \,
\calS\big(\bk, X(x)\big)\,,\eeq
where $\calK(\bk; x)$ is the kernel describing a `non-eikonal' emission, that is,
the emission of a gluon with a generic value of $x$, as computed without any kinematical approximation.
[As the notation suggests, the `eikonal' kernel $\calK(\bk; 0)$ which enters the LO BK evolution is
obtained by letting $x\to 0$ inside the more general kernel $\calK(\bk; x)$.]
\eqn{toyNLON} can be viewed as the sum of Figs.~\ref{fig:eik} and \ref{fig:IF}:
the contribution of the eikonal primary emission cancels out in this sum, so we are left with 
2 diagrams (the bare dipole and a dipole dressed by a non-eikonal gluon emission)
 corresponding to the 2 terms visible in the r.h.s. of \eqn{toyNLON}. 

\comment{
\begin{figure}[t]
\centerline{
\includegraphics[width=.75\textwidth]{dipole_NLO_IF.pdf}}
\caption{\sl }
\label{fig:LOIF}
\end{figure}
}

\begin{figure}[t]
\centerline{
\includegraphics[width=.5\textwidth]{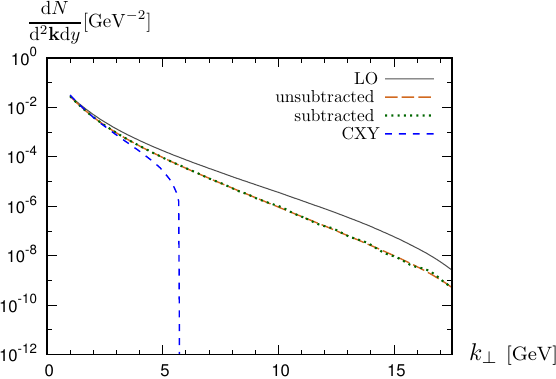}\qquad
\includegraphics[width=.5\textwidth]{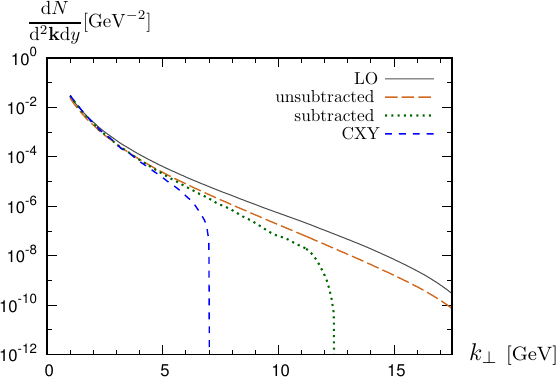}}
\caption{\sl Numerical results comparing the LO result  
in \eqn{LO} with different formulations of the NLO factorization:
``unsubtracted'', cf. \eqn{toyNLON}, ``subtracted'', cf.  \eqn{toyNLOsub},  and
``CXY'', cf.  \eqn{toyCXY}. Left: fixed coupling. Right: Running coupling $\abar(\kt^2)$.
Taken from Ref.~\cite{Ducloue:2017mpb}.}
\label{fig:num}
\end{figure}

On physical grounds, it is {\em a priori} clear that the r.h.s. of  \eqn{toyNLON} must be positive-definite: 
the emission of the primary gluon can only increase the unintegrated gluon distribution of the target.
This is confirmed by the numerical evaluation of \eqn{toyNLON} in  Ref.~\cite{Ducloue:2017mpb},
which yields the ``unsubtracted'' curve in the left panel of Fig.~\ref{fig:num}. However, \eqn{toyNLON} 
is not exactly the same as the hybrid factorization in Ref.~\cite{Chirilli:2012jd}, 
for which the negativity problem occurs. To obtain the latter, one should first separate the LO 
contribution to the cross-section from the NLO 
correction to the impact factor. This can be done by combining Eqs.~\eqref{toyLON}  and
 \eqref{toyNLON} to deduce
\beq\label{toyNLOsub}
 \frac{\rmd N^{pA\to qX}}{\rmd^2\bk\, \rmd \eta}\bigg|_{\nlo}=\,
\calS(\bk,X_g) + \abar\int_{X_g/X_0}^{1} \frac{\rmd x}{x} \,\big[\calK(\bk; x) -
\calK(\bk; 0)\big]\,
\calS\big(\bk, X(x)\big)\,.\eeq
Since obtained via exact 
manipulations, the result in \eqn{toyNLOsub} is identical to that in
\eqref{toyNLON} --- in particular, it is still positive. This is confirmed 
by the numerical results in \cite{Ducloue:2017mpb} (see the ``subtracted'' curve in 
the left panel of Fig.~\ref{fig:num}), which also show that the second term in \eqn{toyNLOsub} --- 
the NLO correction to the impact factor --- is by itself negative, 
but smaller than the LO contribution. The negativity problem only occurs after
additional approximations, which are specific to hybrid factorization and aim at obtaining
a result which looks local in rapidity (here, in $X$): \texttt{(i)} The integral over $x$ in \eqn{toyNLOsub} 
is controlled by $x\sim 1$, hence to the NLO accuracy of interest, one can replace
$X(x)\to X(1)=X_g$ inside the $S$-matrix. \texttt{(ii)} After the previous step, one can extend the integral over $x$ down to $x=0$. We are thus led to the NLO factorization proposed in \cite{Chirilli:2012jd}:
\beq\label{toyCXY}
 \frac{\rmd N^{pA\to qX}}{\rmd^2\bk\, \rmd \eta}\bigg|_{\CXY}=\,
\calS(\bk,X_g) + \abar\int_{0}^{1} \frac{\rmd x}{x} \,\big[\calK(\bk; x) -
\calK(\bk; 0)\big]\,
\calS\big(\bk, X_g\big)\,.\eeq
However, this result suddenly turns out negative when increasing $\kt$, as visible too in the left panel of
Fig.~\ref{fig:num}. As discussed in \cite{Iancu:2016vyg},
this is an artifact of the additional approximations \texttt{(i)},  \texttt{(ii)}, which
artificially enhance the negative contribution of the NLO impact factor [e.g., for any $x\le 1$, one has 
$\calS(\bk,X_g) \ge \calS(\bk,X(x))$].

The LO approximation to the dipole evolution,
cf. \eqn{BKint}, is very bad in practice. A considerably better
approximation, which allows for realistic applications to phenomenology, is obtained by
including the QCD running-coupling (RC) corrections. The respective generalization
of the BK equation \eqref{BKint} is rather straightforward: 
it suffices to replace there $\abar \to \abar(r_{\rm min})$,  where
$r_{\rm min} \equiv \min\big\{|\bx \minus\by|,|\bx \minus\bz|,|\by \minus\bz|\big\}$  
and $\abar(r_{\rm min})$ is the one-loop RC evaluated at $Q^2=4/r_{\rm min}^2$.
But the corresponding generalization of the cross-section for particle production is complicated by
the potential mismatch between the treatment of RC corrections in
transverse coordinate space, and in momentum space, respectively \cite{Iancu:2016vyg}.

Specifically, the NLO cross-section in \eqref{toyNLON} is written in momentum space,
hence the coupling $\alpha_s$ which appears there is naturally interpreted as  $\abar(\kt^2)$
\cite{Iancu:2016vyg}. It may be tempting to perform a similar substitution
in the `subtracted' version, \eqn{toyNLOsub}, but this is not fully right: in the presence of RC corrections,
the second equality in \eqn{toyLON} is only {\em approximately} satisfied\footnote{The momentum-space function
$\calS(\bk,X_g)$ is obtained by first solving the BK equation 
\eqref{BKint} with $\abar \to \abar(r_{\rm min})$ and then performing a Fourier transform; but the effect
of the latter is not faithfully reproduced by replacing $\abar \to \abar(\kt^2)$ in the r.h.s. of
\eqn{toyLON}.}, hence there is some mismatch between the LO quantities which are `added' 
and `subtracted' when going from \eqn{toyNLON} to  \eqn{toyNLOsub}.
Since these quantities are individually
large, such a mismatch may easily lead to unphysical results. And indeed, whereas \eqn{toyNLON}
with $\abar \to \abar(\kt^2)$ is still guaranteed to be positive-definite, this is not the case anymore for
\eqn{toyNLOsub} [with $\abar \to \abar(\kt^2)$], which turns negative at sufficiently large $\kt$,
as illustrated in the right panel of Fig.~\ref{fig:num} (taken again from Ref.~\cite{Ducloue:2017mpb}).

To summarize, our result for the quark multiplicity, as presented (in symbolic notations) in \eqn{toyNLON},
is correct to NLO accuracy and positive-definite. As such it is suitable for applications to the phenomenology.

\bigskip
{\bf Acknowledgments}
The work of E.I. was supported in part by the Agence Nationale de la Recherche project 
 ANR-16-CE31-0019-01. The work of A.H.M.
is supported in part by the U.S. DOE Grant \# DE-FG02-92ER40699.

\bigskip


\end{document}